\documentclass{aa}
\def\lefevre{Le\thinspace F\`evre~}
\def\miralda{Miralda-Escud\`e~}
\def\ms0440{MS\thinspace 0440$+$0204}
\def\etal{et al.~}
\def\approx{$\,\sim\,$}

\def\gequal{$\,\geq\,$}

\def\gthan{$\,>\,$}
\def\lthan{$\,<\,$}

\def\equals{$\,=\,$}
\def\deg{\hbox{$^\circ$}}
\def\sun{\hbox{$\odot$}}

\def\la{\mathrel{\hbox{\rlap{\hbox{\lower4pt\hbox{$\sim$}}}\raise1pt\hbox{$<$}}}}
\def\ga{\mathrel{\hbox{\rlap{\hbox{\lower4pt\hbox{$\sim$}}}\raise1pt\hbox{$>$}}}}

\def\arcmin{\hbox{$^\prime$}}

\def\ltsim{\raise 2pt \hbox {$<$} \kern-1.1em \lower 4pt \hbox {$\sim$}}
\def\ltapprox{\raise 2pt \hbox {$<$} \kern-1.1em \lower 5pt \hbox {$\approx$}}
\def\gtsim{\raise 2pt \hbox {$>$} \kern-1.1em \lower 4pt \hbox {$\sim$}}
\def\gtapprox{\raise 2pt \hbox {$>$} \kern-1.1em \lower 5pt \hbox {$\approx$}}
\def\lefevre{Le\thinspace F\`evre~}
\def\etal{{\it et al.} }

\def\p0{\phantom{0}}

\def\ph1{\phantom{1}}

\begin{document}

   \thesaurus{ (02.07.1) (04.19.1) (11.03.1) (12.03.3) (12.04.1) (12.07.1) (13.25.2) (13.25.3) } 
   \title{A search for gravitational lensing in 38 X-ray \\
         selected clusters of galaxies}
 
   \author{G. A. Luppino\inst{1}, 
I. M. Gioia\inst{1},\inst{2}
F. Hammer\inst{3}, 
O. Le\thinspace F\`evre\inst{4}
\and
J. A. Annis\inst{5}
}
\offprints{I. M. Gioia,  e-mail:gioia@ira.bo.cnr.it, 
Istituto di Radioastronomia del CNR, Via Gobetti 101, I-40129,
Bologna, ITALY}
\institute{Institute for Astronomy, University of Hawaii, 
Honolulu, HI 96822, USA
\and 
Istituto di Radioastronomia del CNR, I-40129 Bologna, ITALY
\and
DAEC, Observatoire de Paris Meudon, 92195 Meudon Principal Cedex, FRANCE
\and
Laboratoire d'Astronomie Spatiale, 13376 Marseille Cedex 12, FRANCE
\and
Experimental Astrophysics Group, Fermilab, Batavia, IL 60510, USA 
}

\date{}

\authorrunning
{Luppino et al.}
\titlerunning
{Gravitational lensing in X-ray clusters}

\maketitle

\begin{abstract}

We present the results of a CCD imaging survey for gravitational
lensing in a sample of 38 X-ray-selected clusters of galaxies.
Our sample consists of the most X-ray luminous 
($L_x$ $\geq$ 2 $\times$ 10$^{44}$ erg\thinspace s$^{-1}$)
clusters  selected from the {\it Einstein Observatory}
Extended Medium Sensitivity Survey (EMSS) that are observable 
from Mauna Kea ($\delta>$ $-$40$^{\rm o}$).
The sample spans a redshift range of 0.15$\leq z\leq$0.823 
and includes 5 clusters with $z>$ 0.5.
CCD images of the clusters were obtained in excellent seeing.
There is evidence of strong gravitational lensing in the form of
giant arcs (length $l\geq8''$, axis ratio $l/w\geq10$) 
in 8 of the 38 clusters. Two  additional clusters contain shorter 
arclets, and 6 more clusters contain candidate arcs
that require follow-up observations to confirm their lensing origin.  
Since the survey does not have a uniform surface brightness limit
we do not draw any conclusion based  on the statistics of the arcs found.
We note, however, that 60\% (3 of 5) of the clusters with
$L_x>$ 10$^{45}$ erg\thinspace s$^{-1}$, and none of the 15
clusters with $L_x<$ 4$\times$10$^{44}$ erg\thinspace s$^{-1}$
contain giant arcs, thereby confirming that high X-ray luminosity 
does identify the most massive systems, and thus X-ray selection 
is the preferred method for finding true, rich clusters at 
intermediate and high redshifts.

The observed geometry of the arcs, most of which are thin, have large 
axis ratios ($l/w > $10), and are aligned orthogonal to the optical major 
axes of the clusters, indicate the cluster core mass density profiles must be 
compact (steeper than isothermal). In several cases, however, there 
is also some evidence, in the form of possible radial arcs, for density 
profiles with finite core radii. 

\keywords{galaxies: clusters: general -- gravitational lensing -- 
X-Rays : galaxies -- Cosmology: observations : dark matter
               }

   \end{abstract}
%

\section{Introduction}

Gravitational lensing by clusters of galaxies is recognized as
a powerful cosmological tool.  By measuring the various properties 
of the gravitationally magnified and 
distorted mirages of background galaxies  (the so-called arcs and 
arclets), we can learn important information about both the lensing 
cluster and the faint, distant,  lensed background galaxies.  
In principle, cluster lenses may provide a means to study normal
galaxies at high redshift ($z\ga 1$); galaxies that would be out of reach 
of even our largest telescopes if not for the magnification of the lensing
clusters. These galaxies are likely to be free from selection effects
since they lie serendipitously behind the foreground clusters and are not 
selected because they are intrinsically bright, or are luminous radio or
X-ray emitters. 
The majority of the arcs and arclets discovered so far
are very blue, and are believed to come from the ubiquitous background
population of faint blue galaxies (FBGs) seen in deep field galaxy
surveys (Tyson 1988; Colless \etal 1990 and 1993; Lilly, Cowie \& Gardner 
1991; Koo and Kron 1992; Lilly \etal 1995 among others).  
It may be possible to use gravitational lensing to constrain the redshift 
range of the FBGs by examining the frequency of lensing as a function of 
cluster redshift (i.e.: Tyson, Valdes \& Wenk, 1990; \miralda 1993b; 
Smail, Ellis and Fitchett, 1994).
But for all its potential as a probe of distant galaxies, cluster 
gravitational lensing is presently finding its most useful application 
as a tool for studying the  clusters themselves.  
Cluster lens systems are relatively common compared to lensed QSOs because 
the high space density of the faint blue galaxies provides a 
convenient background grid whose distortion we can observe around any 
sufficiently massive and concentrated foreground cluster.  
Lensing can be used to measure the
cluster mass, independent of assumptions about virialization of the galaxies
or hydrostatic equilibrium of the hot X-ray-emitting intracluster gas.
Moreover, gravitational lensing provides a means to explore the actual shape
of a cluster's total mass distribution, both visible and dark matter,
at small radii ($r \la 500$\thinspace kpc)
using strong gravitational lensing (giant arcs), and at large radii 
using weak gravitational lensing (\miralda 1991 \& 1992; Kaiser, 1992;
Kaiser \& Squires 1993;  Mellier \etal 1994; Broadhurst, Taylor \& 
Peacock 1994; Smail, Ellis, \& Fitchett, 1994; Kaiser, Squires 
\& Broadhurst 1996; Schneider, 1996; Squires \etal 1996a and 
1996b, Squires \etal 1997; Geiger \& Schneider, 1998; Clowe \etal 1998).

Clearly, a large, homogeneous sample of clusters of galaxies spanning a broad
range of redshift would be extremely valuable.  
Previous attempts to investigate the statistics or frequency of lensing have
relied on optically-selected clusters with all of the problems inherent
to the optical selection (see Lynds \& Petrosian 1989; Smail \etal 1991).
It is well known that projection and contamination problems can be
avoided if clusters are selected based on their primary baryonic
constituent---the hot, intracluster gas that is a copius emitter of 
X-rays---rather than on the optical galaxies which are merely trace components.
We entered this investigation with two primary goals: first, to discover
new cluster lens systems,  and second, to systematically study
a complete sample of distant clusters whose selection criteria were
well defined and designed to avoid the contamination and
superposition problems.  The sample was selected from the {\it Einstein 
Observatory} Extended Medium Sensitivity Survey (EMSS) list of clusters 
given by Gioia \etal (1990a). 

\lefevre \etal (1994) presented preliminary results from an analysis 
of lensing in a subsample of the 16 most X-ray luminous and distant 
EMSS clusters  ($z$\gthan 0.2 and $L_x$\gthan $4 \times 10^{44}$ 
erg\thinspace s$^{-1}$).
Note, however, that this subsample was extracted from the list of
Henry \etal (1992) and {\it did not} include  the 3 added clusters 
with $z$\gthan 0.5 (MS\thinspace 1137+66,  MS\thinspace 0451$-$03, 
MS\thinspace 1054$-$03; Gioia \& Luppino 1994) since their redshifts
were not known at the time. For the most part, the \lefevre \etal
data consisted of $V$ and $I$ ``snapshots'' taken with the CFHT 
that allowed identification of the brightest arcs that might be
present in these clusters.  From this preliminary subsample, they 
concluded that bright arcs with large axis ratios ($l/w >>1$) were fairly
common in X-ray selected clusters, and that this high frequency of 
lensing implied that the mass density profiles of the lensing clusters must be
compact ($r_c < 100$ kpc; see also Hammer \etal 1993; Wu \& Hammer 1993;
Hammer \etal 1997).

In this paper, we present the results from the 38-cluster 
sample. In \S 2, we discuss the properties of the  EMSS galaxy 
cluster sample and our selection criteria for this 
survey.  We describe the optical observations and data reduction and 
calibration in \S 3.  In \S 4, \S 5 and \S 6 we present our data and
describe in detail those clusters that contain giant arcs.
In addition to the obvious lensing cases, we also describe those 
systems with candidate arcs whose lensing origin needs
follow-up confirmation, either with deeper imaging or spectroscopy.
In \S 7, we investigate the lensing fraction as a function of both 
redshift and  X-ray luminosity. We also discuss the constraints the 
EMSS cluster lenses place on the cluster mass density profiles.
Finally, in \S 8, we summarize our results 
and discuss the prospects for future work.

Throughout this paper, we assume a cosmology where 
$H_0$=50\thinspace km~s$^{-1}$~Mpc$^{-1}$, $q_0\,$=$0.5$, but
we occasionally use both conventional parameterizations of the 
Hubble constant, $h_{50} \equiv H_0/50$  and $h \equiv H_0/100$,
depending on the context of the discussion.

\section{The EMSS arc survey cluster sample}

The {\it Einstein Observatory} Medium Sensitivity Survey (EMSS; 
Gioia \etal 1990a, Stocke \etal 1991, Maccacaro \etal 1994) 
has been for many years the only large and sensitive X-ray 
catalog from which an X-ray selected sample of distant clusters 
can be drawn. Consequently, the EMSS cluster sample has been under study 
by a number of groups for a variety of cosmological investigations
(Annis 1994; Gioia \etal 1990b; Henry \etal 1992; Vikhlinin \etal 1998b; 
Henry, 1997; Fahlman \etal 1994; Luppino \& Kaiser 1997; Clowe \etal 1998;
Luppino \& Gioia 1995; Carlberg \etal 1997a, 1997b and 1998; 
Donahue \etal 1998). These investigations range from the study of galaxy 
evolution in  $z\simeq0.3$ X-ray selected clusters, in evolution of the 
X-ray luminosity and of the temperature function  of clusters, in
detection of weak  gravitational lensing and consequent mass estimates, 
to determination  of cluster virial masses and of the cosmological density 
parameter $\Omega_{0}$.  Some of these studies have made X-ray
followed up observations of individual clusters using the{\it ROSAT}
or {\it ASCA} satellites (Donahue \& Stocke 1995; Donahue 1996;
Donahue \etal 1998; Gioia \etal 1998a).

Until the {\it ROSAT} North Ecliptic Pole (NEP) region (Henry
\etal 1995; Mullis, Gioia \& Henry 1998)  of the all-sky survey is 
completely identified, the EMSS is the  only large and sensitive 
X-ray  catalog from which an X-ray  selected sample of distant 
clusters can be drawn. Even if there are several EMSS-style cluster
surveys working from the {\it ROSAT} data archive of pointed observations
(Rosati \etal 1995 and 1998; Scharf \etal 1997; Collins \etal 1997;
Vikhlinin \etal 1998a), most of these  are still works in progress 
and do not cover a large area of sky ($>$ 700 square degrees) as
the EMSS, with the exception of the Vikhlinin \etal sample which covers 
about 160 square degrees, a fourth of the EMSS area.

During the course of this survey, the total number of clusters in the
sample has fluctuated slightly. We added new distant EMSS clusters that
previously had no measured redshift, but were obviously at distances 
$z$\gthan 0.5 and, from their detected X-ray flux, clearly met our 
selection criteria. We also removed sources that, after follow-up 
observations, turned out not to be clusters.  For example, three of the 
sources listed in Gioia and Luppino (1994), MS\thinspace 1209$+$3917, 
MS\thinspace 1333$+$1725  and MS\thinspace 1610$+$6616,
were removed from the cluster list after {\it ROSAT} observations 
revealed the they are unresolved (MS\thinspace 1209$+$39 has been
identified with a Bl Lac object by Rector, Stocke \&  Perlman, 1998, 
MS\thinspace 1333$+$1725 is a star  and MS\thinspace 1610$+$66 is 
still unidentified).

There are presently 100 EMSS sources classified as clusters.
The cluster subsample we chose for the 
arc survey was subjected to the following criteria. 
First, the sources had to lie North of declination
$\delta$\gequal$-40 \deg$ to be observable from Mauna Kea. Second,
the fluxes of the sources in the
$2.4\thinspace \arcmin \times 2.4 \thinspace \arcmin$ IPC detection cell
had to exceed $1.33 \times 10^{-13}$ erg\thinspace cm$^{-2}$\thinspace s$^{-1}$ 
after converting from IPC counting rates with
a thermal spectrum of 6 keV temperature and correcting for
the galactic absorption in the direction of each source, but with 
no IPC point response function correction applied. Third, 
we restricted our sample to clusters with redshifts 
$z$\gequal 0.15, and fourth, we required that the 
X-ray luminosity be greater than 
$L_{\scriptsize 0.3-3.5\,{\rm keV}}$$\,\geq\,$$2.0\times 10^{44}$ 
erg\thinspace s$^{-1}$
to select for deep potential wells which are most likely to 
exhibit gravitational lensing. These criteria resulted in
38 clusters spanning a large  redshift range ($0.15 \leq z \leq 0.823$) 
and an order of  magnitude in X-ray luminosity 
($2\times 10^{44}\; {\rm erg s}^{-1} \leq L_x \leq 2\times 10^{45}\; 
{\rm erg s}^{-1}$). For each of the 38 clusters optical, radio and 
X-ray data can be found in Gioia \& Luppino (1994), including wide
field CCD images  ($>$1\thinspace Mpc$\times$1\thinspace Mpc in 
the cluster frames).

Although we used X-ray selection to avoid obvious optical selection 
effects, it would not be fair to claim that the EMSS cluster sample 
is completely free from selection biases. As pointed out by Donahue,
Stocke \& Gioia (1992; hereafter DSG), the EMSS is not, strictly speaking, 
a flux limited sample. Instead, the detection of EMSS sources is limited 
by central surface  brightness.  Consequently, there may be a bias 
toward clusters with centrally peaked X-ray surface
brightness since the EMSS detection algorithm was optimized for
detecting point sources in the IPC  $2'.4 \times 2'.4$ detection cell. 
Clusters with more extended emission at lower surface brightness would
have been resolved by the detector and may have been missed by failing
to meet the minimum detection criterion in the central cell
(note that this problem is more severe at lower redshifts).
One might argue that the EMSS might preferentially select cooling flow
clusters (Pesce \etal 1990; Edge \etal 1992) given the possible bias toward 
centrally peaked objects.  DSG, however, make compelling arguments for why
EMSS clusters cannot all be cooling flow clusters; namely that there do exist
non-cooling flow clusters with large $L_x$ and small core radii. 
Moreover, clusters with large core radii tend to have lower $L_x$ and thus
would be excluded from the EMSS sample for that reason.  
Additional evidence that the EMSS does not miss clusters which are 
not cooling flow clusters is given by the agreement between the X-ray 
luminosity function of the EMSS clusters with 
$z$\equals 0.14$\,\rightarrow\,$0.2, and the X-ray luminosity function 
derived by earlier studies (Piccinotti \etal 1982) which use large-beam, 
non-imaging detector fluxes or with the X-ray luminosity function
of the {\it ROSAT} Brightest Cluster Sample  (Ebeling \etal 1997).
The agreement indicates that this bias, if present in our sample, 
is small, or  at least it is at work in the same manner in the 
non-imaging and in the {\it ROSAT} data.

\section{Observations}

Imaging observations of the clusters were carried out during the period from
May of 1992 through January of 1994 using the University of
Hawaii (UH) 2.2m telescope equipped with a Tektronix
2048 $\times$ 2048 pixel thinned, back-illuminated,
anti-reflection-coated CCD mounted at the f/10 Cassegrain
focus. The image scale was $0''.22$/24$\mu$m-pixel and the field of view
was $7'.5 \times 7'.5$ ($1.54h_{50}^{-1}$\thinspace Mpc
$\times$ $1.54h_{50}^{-1}$\thinspace Mpc at
$z$=0.15 and $3.74h_{50}^{-1}$\thinspace Mpc $\times$ 
$3.74h_{50}^{-1}$\thinspace Mpc at $z$=0.82). 
$R$ band images of all, and $B$ band images of most of the
clusters were acquired using this configuration. 
We also took additional $V$ and $I$ images of a few selected
clusters using both the UH 2.2\thinspace m and HRCam (McClure \etal
1989) on the CFHT.  
In total, there were 8 observing runs: May, June,
September and October of 1992, February, May and November of 1993, and 
January of 1994.  The majority of the data are of excellent quality. 
Except for the February 1993 run, the images were taken in photometric 
conditions and in good seeing.
The $R$ seeing ranged from $0''.5$ to $1''.3$ FWHM with a median
value of $0''.8$ FWHM. The seeing for the $B$ images was slightly
worse with a median  of $0''.9$ FWHM.

We chose the $B$ and $R$ bandpasses for the following reasons.
The $R$ bandpass is ideal for observing the distant clusters
in the redshift range spanned by the EMSS sample.  Furthermore, the center
wavelength of the $R$ filter occurs near the peak of the CCD quantum
efficiency, and the night sky is darker than in a redder bandpass such as $I$.
Although gravitationally lensed arcs are known to be relatively blue, they 
are still easily detected in the $R$ band.
The additional $B$ or $V$ band images were taken to  allow us to recognize
any gravitational arcs by their relative blue color compared to the
red cluster galaxies.

In order to build up our long exposures, we used the standard
``shift-and-stare'' or ``dithering'' technique where we took a number of short
integrations (typically 600\thinspace s or 900\thinspace s each) with
the telescope shifted $\sim$\thinspace 20--30$''$
between exposures, allowing us to assemble a
median-filtered stack of the disregistered images to use for
flattening the data. Typically, the set of all images in a single color
spanning an entire night, and often an entire run ($\sim$50--100 frames),
were used to build the median skyflat.
Each separate cluster image was first bias subtracted and then flattened with
the normalized median flat.  The final image was then produced by
shifting the individual images into registration (integer pixel shifts)
and adding them while cleaning them of cosmic ray hits.

\section{Properties of the clusters and arcs}

In this section, we summarize the optical, X-ray and radio data for
the 38 EMSS clusters, and describe in detail those sources classified 
as gravitational lens systems.  Updated
optical coordinates for these clusters can be found in 
Gioia \& Luppino (1994).
The general properties of the clusters are presented in Table 1.
Columns (1) and (2) list the name and redshift of each EMSS cluster.
Columns (3), (4) and (5) contain the X-ray properties of the cluster: the
detected flux in the 
$2.4\thinspace \arcmin \times 2.4 \thinspace \arcmin$ IPC 
cell ($F_x^{\rm Det}$) and total flux ($F_x^{\rm Tot}$) in 
erg~cm$^{-2}$~s$^{-1}$, and the corrected cluster X-ray 
luminosity ($L_x$) in the 0.3$\,-\,$3.5 keV band in erg\thinspace s$^{-1}$ 
respectively. The total flux and X-ray luminosity have been corrected  
following the prescription outlined by Gioia \etal (1990b) and  
Henry \etal (1992), which accounts for any extended cluster emission 
that may have been present outside the IPC detect cell. This correction 
can be substantial at low redshift 
($z$\lthan 0.2) but is small at high  redshift.

\begin{figure*}
\vskip 9.0truein
\includegraphics{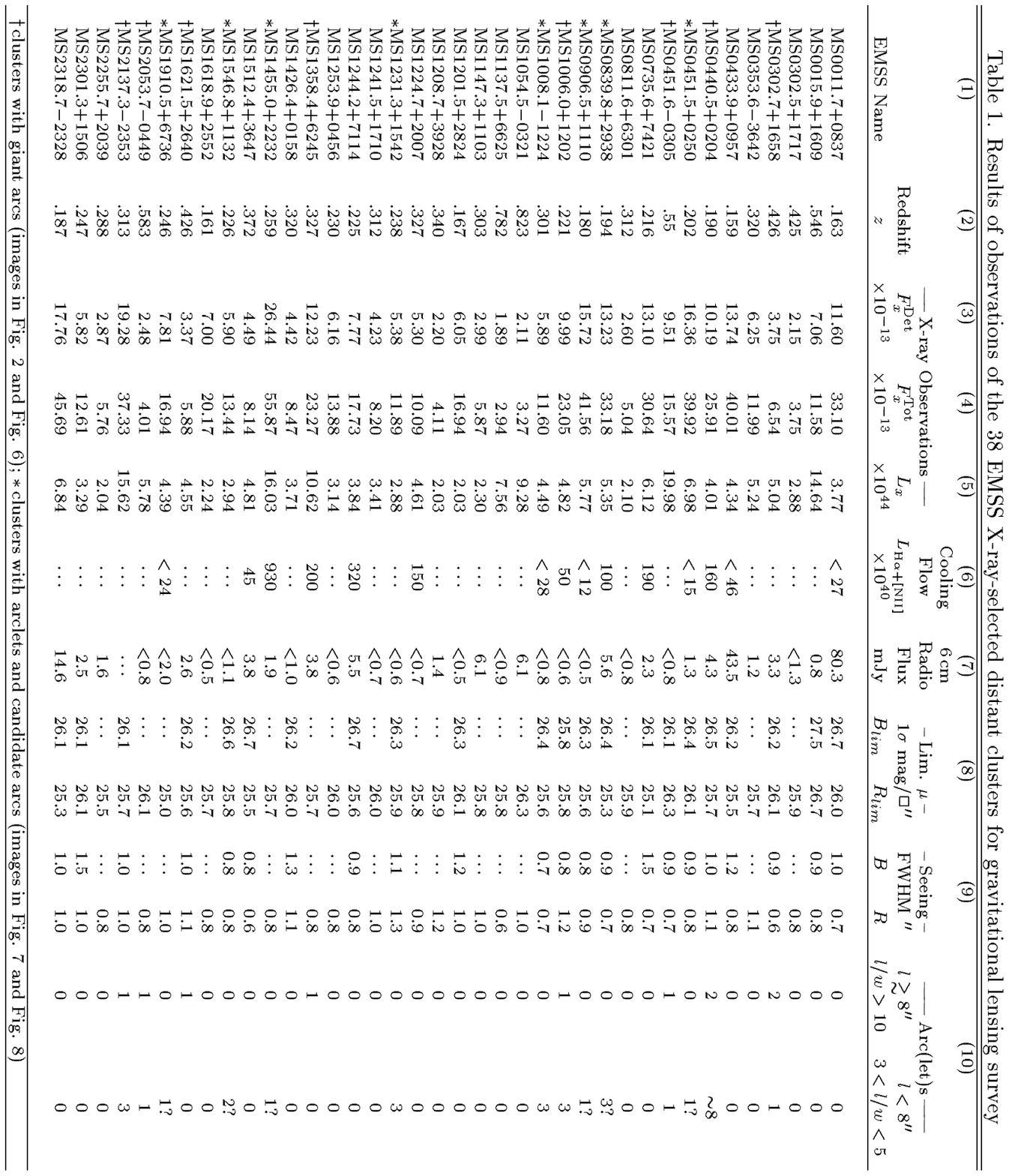}
\end{figure*}
Column (6) lists the H$_\alpha$$\,+\,$[NII] luminosities 
(in erg\thinspace s$^{-1}$)
taken from Donahue, Stocke \& Gioia (1992) who interpreted this line
emission as evidence for a cooling flow.
Not all of the clusters in our sample were observed by DSG, so 
only those clusters 
with an entry in this column have cooling flow measurements. 
Column (7) contains the 6\thinspace cm radio flux
measured with the $VLA$ (Gioia \etal 1983; Stocke \etal 1991;
Gioia \& Luppino 1994). Columns (8) and (9) indicate the 
quality of the optical data on each cluster by 
listing the $B$ and $R$ limiting magnitudes ($1\sigma$) and the seeing.  
We obtained $R$ images for all 38 of the clusters, but $B$ images for
only 22.  For many of the clusters, however, we do have $V$ and/or $I$
images, and in several cases, these are quite deep.  
Finally, in column (10), we list the number of arcs ($l\ga 8''$) and 
arclets ($l < 8''$) seen in each cluster.  
The ($\dagger$) and ($\ast$) symbols that precede some of the cluster
names indicate those clusters that have giant arcs or
candidate arcs and appear in Figures 2 and 6, and in Figures 7 and 8,
respectively. For the purposes of this investigation, we define a giant 
arc as a large, elongated structure having a length $l$\gthan 8$''$ 
and an axis ratio $l/w>10$.
We consider both high surface brightness (``luminous'') arcs and fainter,
low surface brightness ones as well.  Arclets are simply short arcs
($l$\lthan 8$''$) with axis ratios 3\lthan$l/w$\lthan 10.
The clusters in Table 1 have considerable variation in their optical
appearance, ranging from poor, compact groups to rich regular and irregular
clusters. Cases of multiple nucleus cD galaxies, and clusters with optical 
substructure exist.  Eventually, a quantitative morphological 
classification (e.g. Bautz-Morgan) would be desirable, but such a 
classification requires galaxy photometry and often spectroscopy (to
verify the cluster membership of the brightest galaxies) and is 
beyond the scope of this paper. We refer the reader to the $R$ band 
optical images published in Gioia  \& Luppino, 1994.

\section{Clusters with giant arcs}

Eight of the 38 EMSS clusters contain giant arcs. 
To extract quantitative information from cluster gravitational 
lenses, one needs a large, homogeneous, well-selected sample of clusters
for which there exist measurements of the lengths, widths, radii
of curvature, positions and orientations of any gravitational images.  
Arc lengths and widths are related to the size and ellipticity
of the sources, while arc widths are also strongly affected by
the steepness of the cluster mass density profile.  The radii of 
curvature and orientation of the arcs depend primarily on the 
potential of the clusters, providing important clues about the
density profile and mass substructure.

We note that making these measurements and attempting to force
the often complex lensing configurations into a simplified geometry
is difficult and problematic. We choose to list the arc location
with respect to the brightest cluster member, which is often
not the true center of mass. Measuring the arc length, width, and
especially the radius of curvature is more difficult. The arcs often
have structure along their length and variation in their width, and are
usually too short for an accurate determination of their radius of
curvature, $R_c$. In spite of these difficulties, we at least attempt to
estimate these quantities and list them in the tables for future use.
We measure the arc total magnitudes by integrating all the flux from
the arc to the point where the lowest isophote blends into the local
sky background.  In some cases where the arc is embedded in the halo
of one or more galaxies, the flux from the arc is contaminated
by light from this halo, and we attempt to subtract the halo
flux in order to measure the true arc magnitude.

In Table 2 we list the observed properties of all the arcs in the
eight clusters that contain at least one giant arc. Refer to the 
diagram in Figure 1 where we illustrate the various arc geometry 
terms. Columns (1), (2), and (3) contain the cluster name,
the cluster (lens) redshift ($z_l$), and the arc name or label 
respectively.
The length, $l$, seeing-deconvolved width, $w$, axis ratio $l/w$, and 
radius of curvature $R_c$, are listed in columns (4), (5), (6) and (7).
The radius of curvature can be determined
by measuring the chord length across the two ends of the arc, and the
sagittal depth from the arc to this chord, assuming the arc is circular.
Column (8) contains the distance, $d_c$, from the (approximate) center
of the arc to the center of the cluster.  Often the 
true cluster center is ill defined, so for our purposes, we take the
position of the optically dominant galaxy as the cluster center unless
otherwise specified. The angle $\theta$ 
listed in column (9) refers to the angular position of the center
of the arc with respect
to the center of the cluster, where 0$^{\rm o}$ is defined as north
and positive rotations are measured in the counterclockwise direction,
while the angle $\varphi$, listed in column (10), is a measure of the
orientation of the arc with respect to the cluster radius vector (again
the ccw direction is positive). Column (11) contains the (approximate) 
position of the center of curvature ($x_c$, $y_c$)  with respect to 
the cluster center. These numbers are given in arcsec with E and N being 
the positive directions. Columns (12), (13) and (14) list the $B$ and $R$ 
magnitudes and $(B-R)$ colors of the arcs (note, $V$ and $I$ magnitudes 
for some of the arcs can also be found in \lefevre \etal 1994). Finally,
column (15) contains references to more detailed studies for each cluster.

\begin{figure*}[t]
\vspace{5.1in}
\includegraphics{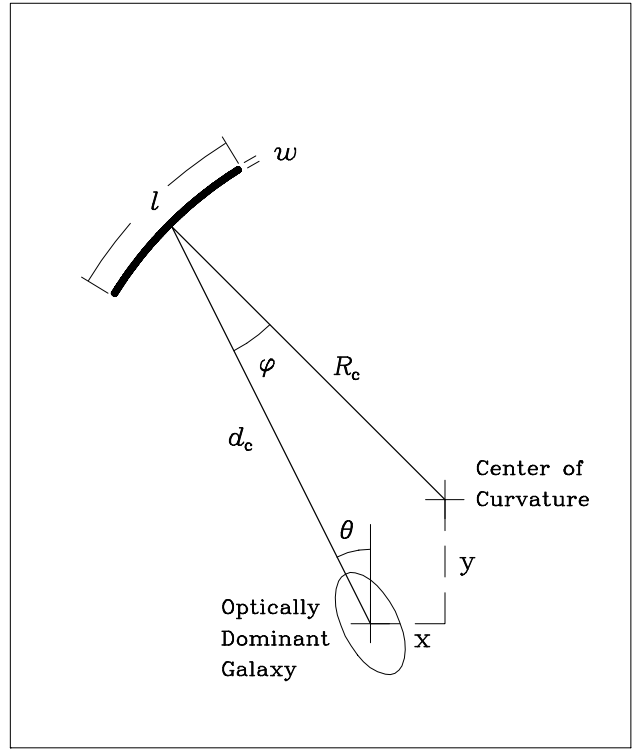}
\includegraphics{GioiaI_fig1_labels.ps}
\vspace{-1.1in}
\begin{center}
\caption{Definition of arc geometry parameters.
}
\end{center}
\end{figure*}

\begin{figure*}[tbhp]
\vskip 9.0truein
\includegraphics{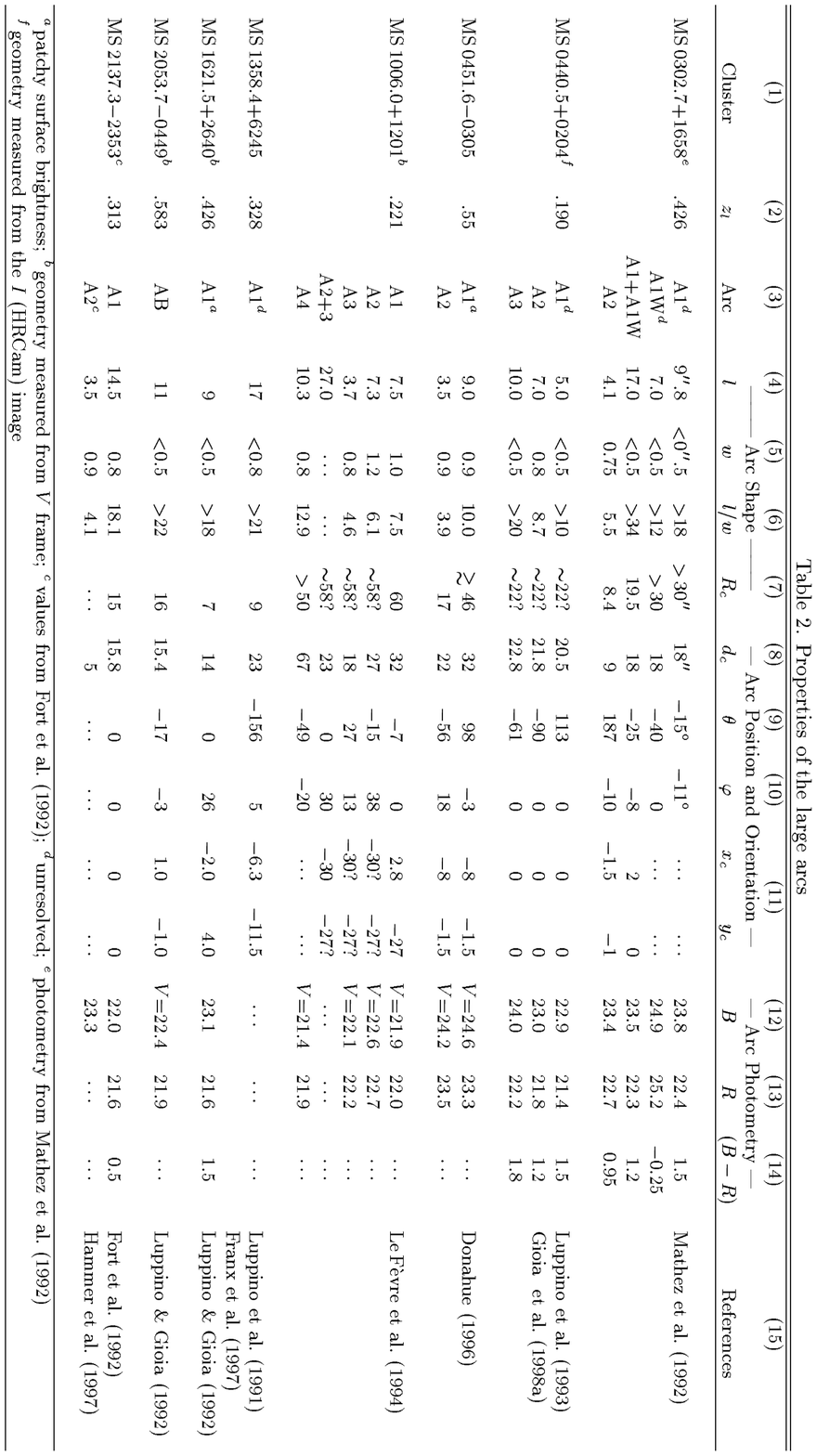}
\end{figure*}

In the following subsections, we describe each giant arc system in detail.

\vskip 0.5cm

\subsection{MS\thinspace 0302.7$+$1658}

MS\thinspace 0302.7$+$1658 at $z$=0.426 was the first EMSS cluster 
found to contain a lensed arc (see Figure 2).  
The arc is first mentioned by Giraud (1991).
Mathez \etal (1992) presented detailed observations of the cluster and 3
arcs: the long ($l\simeq 10''$), so-called ``straight''  arc A1 located between 
the two bright elliptical galaxies G1 and G2 (in the Mathez \etal 
labelling); a shorter arc A2 ($l$\approx 8$''$) to the 
south of the brightest cluster galaxy G1; 
and a faint arc A1W attached to the western end 
of A1.  Giraud (1992) also showed limited evidence for variability in two
additional arclets seen in images taken in 1989 and 1991.
Models by Mathez \etal reproduce the geometry and shape of the arcs for 
assumed redshifts of $z$\approx 0.8 and $z$\approx 0.6 for 
A1 and A2 respectively.

In our $R$-band CCD images taken in $0''.6$ seeing 
we can easily see the 3 large arcs. We do not see the arclets mentioned
by Giraud (1992). In Figure 3, we also show a true color
image of the cluster core where we used the $I$, $R$ and $V$ frames
for the rgb colors.  The giant arc between galaxies G1 and G2 
(a combination of A1 and A1W) is over $20''$ long and 
appears to be circular centered on G1.  However upon closer inspection, 
we verify that the western and eastern components of A1 are both 
more-or-less straight and are folded at the point where they join
to appear concave. Mathez \etal modeled the cluster potential as a 
bimodal configuration with the main mass deflector located near the 
position of the brightest cluster galaxy (BCG) and
the secondary deflector located near the second ranked galaxy to the
NW.  This model is the simplest way to produce  ``straight'' arc images
as in the example of A2390 (Pello \etal 1991; Pierre \etal 1996). 

Fabricant, Bautz \&  McClintock (1994) have obtained spectra of a number of
galaxies in MS\thinspace 0302$+$1658 (as well as the neighboring cluster
MS\thinspace 0302+1717) and find the cluster has a velocity dispersion
of 921$^{+192}_{-123}$ km\thinspace s$^{-1}$.  They also observe 
photometric and spectroscopic evidence of galaxy evolution (i.e. the
Butcher-Oemler effect) with 20--30\% of the galaxies having strong 
Balmer absorption and emission lines indicative of a recent episode 
of star formation. Carlberg \etal (1996) measure a lower value for 
the velocity dispersion of 646$\pm$93 km s$^{-1}$ using 27 galaxies.
Ellingson \etal (1997) have produced a photometric and 
spectroscopic catalog for the galaxies in MS\thinspace 0302$+$1658 
bringing the total number of cluster members with measured redshifts 
to 94, but a new value for the velocity dispersion has 
not been published yet.

Kaiser \etal (1998) have performed a weak lensing study 
of the supercluster at 0.4 using deep $I$ and $V$ band 
images taken with the UH8K mosaic camera at the CFHT. The 
supercluster is composed by the three clusters MS\thinspace 0302.7$+$1658 
at z$=$0.426, MS\thinspace 0302.5$+$1717 at z$=$0.425
(both X-ray selected) and of the optically selected cluster 
CL\thinspace 0303$+$17, at z$=$0.418, discovered by Dressler 
\& Gunn  in 1992. All of the major concentrations apparent in the X-ray 
and optical images are detected in the mass reconstructions, and
indicate that most of the super-cluster mass, like the
early type galaxies, is concentrated in the three X-ray clusters.
A mean mass to light ratio for the clusters of 
M/L$_{B}$ $\simeq$ 260h is obtained. The implication of the results 
for the cosmological density parameter is also discussed.

\subsection{MS\thinspace 0440.5$+$0204}

MS\thinspace 0440$+$0204 at $z$=0.190 is an optically poor cluster with
a compact, multiple-nucleus cD galaxy surrounded by a large halo in which
are embedded a number of blue, circular structures that appear to be 
lensed arcs and arclets (Luppino \etal 1993).  The largest arc has
a length $l\simeq 10''$ and remains unresolved in $0''.5$ seeing. Most of
the arcs and arclets lie on or near a 24$''$ (100\thinspace $h_{50}^{-1}$
kpc) radius critical line.  Luppino \etal computed an enclosed mass and 
central mass-to-light ratio of $1.0\times 10^{14} M_{\sun}$ and
110 $M_{\sun}/L_{\sun}$ respectively  assuming a source redshift 
of $z_s\simeq 0.4$. This cluster has been studied 
in detail by Gioia \etal (1998a) who present a combined analysis 
of X-ray imaging and spectroscopic data and HST data.  From possible
multiple images formed by gravitational lensing of five background
sources, Gioia \etal ~ derive limits to the mass distribution  in the 
range 50$-$100 $h_{50}^{-1}$ kpc. For the central 
100\thinspace $h_{50}^{-1}$ kpc region, the possible range in 
projected mass is 6.6$-$9.5$\times10^{13}$ $h_{50}^{-1}$ M\sun. 
At about 600 kpc from the center of the cluster  a simple $\beta$ model 
fit to the X-ray data yields a mass of (1.3$\pm$0.2)$\times$10$^{14}$ M\sun.
At 100 kpc, the lower limit mass from lensing is a factor 2 greater than
the X-ray determined mass. In order to reconcile
the different mass estimates, Gioia \etal ~tentatively explore a model 
where the mass profile increases more rapidly than the X-ray $\beta$ model 
at large radii.

\subsection{MS\thinspace 0451.6$-$0305}

MS\thinspace 0451.6$-$0305 is a spectacular example of a massive,
X-ray-luminous,  cluster of galaxies.
The cluster has $z$=0.55 and is therefore exceptionally 
luminous with $L_x=2.0\times 10^{45}$ erg\thinspace s$^{-1}$); the most luminous
cluster in the EMSS and among the most luminous cluster X-ray source known.

The high X-ray luminosity is confirmed by Donahue \& Stocke (1995) 
who obtained a 16 ksec {\it ROSAT} PSPC image of MS\thinspace 0451$-$0305.  
They also see a slight elongation
of the X-ray isophotes with roughly the same EW orientation as the 
optical galaxies. Donahue (1996) has also  obtained {\it ASCA} data
that yield a high value for the  temperature of 10.4$\pm$1.2 keV.
Combining the temperature of the gas with the image parameters of the 
{\it ROSAT} PSPC, Donahue obtains a total mass within a radius of 
1 h$^{-1}_{50}$ Mpc
of 9.7$^{+3.8}_{-2.2} \times10^{14}$ h$^{-1}_{50}$ M\sun. This value 
is in agreement with the masses implied by both the virial estimates
using the velocity dispersion of 1371 km s$^{-1}$ reported by the 
CNOC group (Carlberg \etal ~1996, see also Ellingson \etal 1998, for a 
spectrophotometric catalog of the cluster galaxies) and with the 
weak lensing results by Clowe (1998).

Our deep CCD images reveal a rich cluster with a giant
arc and at least one smaller arclet (see Figures 2, 4 and 5). 
A true color image is  shown in Figure 4.  This image is a 
750 $\times$ 750 pixel subarray extracted
from the much larger 2048$^2$ CCD frames, and
was produced using the $I$, $R$, and $V$ CCD images (UH 88-inch)
for the rgb colors. The integration times and seeing (FWHM) in the 
3 colors were 4200\thinspace s and $0''.7$ in $I$, 7200\thinspace s 
and $0''.6$ in $R$, and 8400\thinspace s and $0''.7$ in $V$. The color 
image measures $2'.75 \times 2'.75$ corresponding to 1.2 Mpc $\times$
1.2 Mpc in the cluster frame
(7.39 kpc/arcsec at $z$=0.55 for $H_0$=50, $q_0\,$=$\,{1\over 2}$).
The cluster has an obviously flattened (EW) morphology with red
galaxies that are easily distinguished from the blue field population.
Note that the core of the cluster appears to contain 2 bright galaxies, but 
the true color image reveals that the southern bright galaxy
is quite blue and appears to be a foreground, face-on spiral. 

The high contrast image in Figure 5 is the sum of the $V$, $R$ and $I$
frames and is displayed to reveal the details of the giant arc, A1 and
the bright arclet A2.  
Arc A1 is \approx 9$''$ long and is located $d_c$=32$''$ E of the
brightest cluster galaxy.  It appears to be marginally 
resolved, has variable surface brightness, and at least 1 break along 
its length.  Upon close  inspection, 
A1 appears to be another example of a ``straight'' arc, but precise
measurement of the radius of curvature is difficult.  We find a center-to-edge
deviation from linearity of $\la$1 pixel over the \approx 41 pixel length
of the arc, allowing us to place a lower limit 
on the radius of curvature of
$R_c\ga46''\,$---\thinspace only slightly larger than the distance to the
BCG.  Arclet A2 is \approx 3$''$.5 long, and is located 22$''$ NW of the
BCG.  Lines drawn normal to this arclet and normal to arc A1 intersect at
a point \approx 7$''$.5 W of the BCG. This point is \approx 40$''$
from A1 and is more consistent with the $R_c$ measured for A1.
A2 also appears to be curved and its approximate center is consistent
with this location.

In many ways, arc A1  resembles the  original ``straight'' arc in the
$z$=0.231 cluster A2390 (Pello \etal 1991; Pierre \etal 1996). 
The A2390 arc is resolved, has breaks along its length,
and has a spectroscopically-measured redshift of $z$=0.913.
Both the MS\thinspace 0451$-$0305 arc A1 and the A2390
arc are orthogonal to the optical and X-ray 
major axes of their respective clusters (i.e. the arcs trace 
ellipses aligned with the optical and/or X-ray major axes).
As we mentioned above, the lower limit on $R_c$ for A1
in MS\thinspace 0451$-$03 is not significantly larger than the distance 
to the BCG and is reasonably consistent with the intersection point 
of the normals of A1 and A2.  However, if $R_c>>d_c$, then, as in the 
case of  A2390, we require a secondary mass concentration to the East 
of the cluster where there is no excess of optical galaxies,
suggesting the presence of clumped dark matter.

However, a possible alternative way to produce nearly straight,
elongated images is with a ``marginal'' lens (Kovner 1987b), i.e.
a cluster with a relatively large core radius. In such cases, we would 
expect to see larger arc widths than with the small core
radii clusters, whose arcs would be thin.  
Since the large arc in MS\thinspace 0451$-$03 is marginally
resolved, we must consider this situation seriously. Note, however, that
the orientation of the arc, with respect to both the optical and
X-ray major axes, supports our initial statement that the mass
density profile is compact, and thus our interpretation that
the shape of arc A1 implies mass substructure is the preferred
one.

MS\thinspace 0451.6$-$0305 is in the list of clusters detected by Doug Clowe
(1998) using weak gravitational lensing. The mass distribution created with 
the Kaiser \& Squires algorithm (1993) is centered on the bright central 
galaxy, but the broad EW structure evident in the galaxy distribution
is absent. Instead a moderately large northern extension from the 
central peak present in the mass distribution is consistent with a 
(much weaker) structure seen in the cluster galaxy distribution.

\subsection{MS\thinspace 1006.0$+$1202}

MS1006$+$1202 is a moderately luminous ($L_x=4.8\times 10^{44}$ 
erg\thinspace s$^{-1}$)
cluster at $z$=0.221 with a flattened (NS), irregular 
morphology and an optical core containing
a number of overlapped galaxy images, but with a central galaxy that
has neither the size nor brightness to be called a cD.  
DSG have detected H$_\alpha$ and [NII] emission suggesting the presence
of a cooling flow. Carlberg \etal 1996, measure a velocity dispersion of 
906 km s$^{-1}$ from 26 cluster members.

This cluster contains a remarkable assortment of faint, blue,
linear structures that appear to be lensed images (\lefevre \etal 1994).
There are three arclets, A1, A2, and A3, located \approx 22$''$--30$''$
to the North of the BCG. 
All three arclets are oriented roughly perpendicular to the
major axis of the optical galaxy distribution (see the image in
Figure 2, this paper, and in Figure 1a in \lefevre et\thinspace al.).  
The arclets have lengths ranging 
from $l$\approx 3$''$.7 for A3 to $l$\approx 7$''$.5 for A1. All three
appear to be marginally resolved and have no apparent curvature,
but they are so short that any slight curvature would be difficult to
recognize. Furthermore, some portions of these arclets may be obscured
by a bright ($R$\approx 15) star located near the arclet positions.
Presumably arclets A2 and A3 could be 2 pieces of the same large
arc since lines extending perpendicular from the centers of
these arclets intersect at the same point \approx 58$''$ to the SW,
and the break between the two arclets occurs close to the foreground
star. The length of A2 and A3 combined is \approx 27$''$.
The longest single arc-like structure, A4,
is located 67$''$ NW of the BCG and is \approx 10$''$ long. 
There is some question whether this particular object is
truly a gravitational lens image.  The arc appears to be linear, 
with galaxies apparently at both ends, and might possibly be some
kind of tidal tail from an interacting system.
If A4 is a gravitational image, then this cluster may have a 
peculiar underlying mass distribution. The absence of curvature for 
A4 suggests the presence of a secondary deflector to the West, where 
there is no obvious excess of optical galaxies (see the wide field
CCD image in Gioia \& Luppino 1994).
Furthermore, the orientation of A4 is odd when compared 
to the more-or-less orthogonal orientation of A1--3, although without
detailed modelling, it is difficult to say anything quantitative at
this time. 
Again, we note that an alternative interpretation is to
consider this cluster an example of a marginal lens with a large
core radius, rather than considering the absence of curvature in the
arcs as evidence for invisible mass substructure.

\subsection{MS\thinspace 1358.4$+$6245}

MS\thinspace 1358$+$6245 is a very rich cluster at $z$=0.328 and is
extremely X-ray luminous with $L_x=1.1\times 10^{45}$ erg\thinspace s$^{-1}$.
This cluster has been the focus of a number of investigations of
galaxy evolution in distant, X-ray selected clusters (Luppino \etal
1991; Fabricant, McClintock \& Bautz 1991; Annis 1994; Kelson \etal 1997), and 
displays the Butcher-Oemler effect with a photometrically
and spectroscopically determined blue galaxy fraction of
$f_b\sim20\%$.  During the course of these various investigations,
lensed arcs were looked for but not seen, probably because
the typical seeing in these data was of order 1.3$''$--1.5$''$.
From WFPC2 images obtained with HST, Franx \etal (1997) 
discovered serendipitously an extremely red arc. Keck II spectra of the
arc revealed to be the lensed image of a galaxy at z$=$4.92, among the 
most distant galaxies known. Both Yee \etal 1998
and Fisher \etal 1998, have performed  a spectroscopic 
survey of this rich cluster. Yee \etal find evidence for a second 
cluster at the same redshift as MS\thinspace 1358$+$6245, located at 
the southern edge of the central cluster field.
From their catalog of 232 cluster members Fisher \etal 1998
derive a mean redshift of z$=$0.3283$\pm$0.0003 and a velocity 
dispersion of 1027$^{+51}_{-45}$ km s$^{-1}$ in fair
agreement the value of 987$\pm$54 determined by Carlberg, Yee, 
\& Ellingson 1997. Fisher \etal (1998) show that there  is significant 
evidence for substructure in the central  part of the cluster and 
that the distribution of line-of-sight velocities  departs significantly 
from a Gaussian. They identify two subgroups with at least 10$-$20 
members and dispersions of \lthan 400 km s$^{-1}$. Note that  they
concentrate on the main body of the cluster and that the second
concentration found by Yee \etal south of the cluster center is outside 
the Fisher \etal field limits. The presence of two subclumps in the cluster,
implies that MS\thinspace 1358$+$6245 has not yet
reached equilibrium and is still in the process of virialization 
and accretion of new members. 
van Dokkum \etal (1998)  investigate the 
color-magnitude relation for MS\thinspace 1358$+$6245 
using a wide-field mosaic of multicolor  WFPC2 images. 
A weak lensing study by Hoekstra \etal 1998, finds a total projected
mass within 1 Mpc h$^{-1}_{50}$ of (4.4$\pm$0.6)$\times10^{14}$M\sun ~and 
a M/L ratio =(90$\pm$13)h$_{50}$ M\sun/L$_{V}$\sun ~consistent with 
being constant with radius. These authors use the maximum probability 
extension of the original Kaiser \& Squires (1993) 
algorithm and compare the resulting mass map to the result from a finite 
field construction algorithm developed by Seitz \& Schneider 
(1996 and 1998). MS\thinspace 1358$+$6245 was also studied 
with {\it ROSAT} and {\it ASCA} by Bautz \etal 1997, and by Allen (1998).
The results by Allen, who takes into account the effects of cooling 
flows on the X-ray images and spectra, imply a projected mass of 
4.2$^{+4.1}_{-0.8}$$\times10^{14}$ M\sun ~in excellent agreement with
the weak lensing mass by Hoekstra \etal 1998. 

As part of this survey, we obtained $R$-band images in $0''.8$ seeing
with the UH 2.2\thinspace m telescope and $V$-band images in $0''.6$
seeing with HRCam on the CFHT.  Although there are no bright
arcs visible, we did find a large, very faint, curved arc located to the SW
of the central galaxies, which corresponds to the red arc 
serendipitously discovered by Franx \etal 1997.  The arc is barely 
discernable when the $R$-band image is displayed with ``normal''
contrast at the top of Figure 6, but can be seen clearly
in the high contrast image in the lower panel. 
The same faint, curved structure is also present in our
$V$-band CFHT images. The arc is located 23$''$ (133$\,h_{50}^{-1}$ kpc) 
from the BCG and is 17$''$ long with a small
radius of curvature of $R_c\simeq 9''$ (75$\,h_{50}^{-1}$ kpc). 
The center of curvature is displaced 13$''$ SW of the BCG.

\subsection{MS\thinspace 1621.5$+$2640}

MS1621$+$2640 is an optically rich cluster at $z$=0.426
with one large, faint arc located close to the second brightest cluster
galaxy  (Luppino \& Gioia 1992). The cluster X-ray luminosity is 
$L_x=4.6\times 10^{44}$ erg\thinspace s$^{-1}$. Note the $L_x$ listed
in the original Luppino \& Gioia (1992) paper assumed $q_0=0$ and
used a point source correction rather than the cluster extended 
emission correction as applied later by Henry \etal (1992) and 
Gioia \& Luppino (1994), and as used in this survey.
Ellingson \etal (1997) published photometric and redshift catalogs of 
galaxies in the field of the cluster as part of the CNOC 
cluster redshift survey.

The arc is \approx 9$''$ long, has a patchy surface brightness,
and has a very small radius of curvature
($R_c\simeq 7''$).  The arc is located \approx 14$''$ from the 
BCG and the center of curvature appears to be located \approx 3$''$
S of galaxy 2, between it and the BCG.  

\subsection{MS\thinspace 2053.7$-$0449}

MS\thinspace 2053$-$0449 has a redshift of $z$=0.583 and an X-ray 
luminosity 5.78$\times 10^{44}$ erg\thinspace s$^{-1}$.
It is not very optically rich, especially when compared with 
its other $z$\gthan 0.5 EMSS counterparts like 
MS\thinspace 0016+1609 or MS\thinspace 0451$-$0305.
Luppino \& Gioia (1992) presented high resolution ($0''.6$ seeing) 
$V$ band images of MS\thinspace 2053$-$0449 taken with HRCam on the CFHT
that revealed a large arc-like structure located \approx 16$''$ from the 
BCG.  The arc is \approx 11$''$ long and breaks into 2 distinct clumps, 
labeled A and B, with a  center of curvature that appears to be close 
to the position of the BCG. Luppino \& Gioia also noted a very 
faint arclet, labeled C, located closer in to the BCG.

Clowe (1998) presents a weak lensing
study for this cluster which is the least massive of the z\approx0.55
clusters in the EMSS. The mass profile, generated by aperture densitometry is
well fit by a ``universal'' CDM profile (Navarro, Frenk \& White, 1996)
with parameters r$_{200}$=520 h$^{-1}$ kpc and c$=$2  assuming a background 
galaxy redshift z$_{bg}$=1.75. MS\thinspace 2053$-$0449 is also well fit
by an isothermal sphere model with a velocity dispersion of $\sigma=$700 
km s$^{-1}$ for z$_{bg}$=1.75, indicating that the cluster is close
to virialization.

Kelson \etal 1997, measure structural parameters and central velocity 
dispersions for the galaxies in MS\thinspace 2053$-$0449  to define the
fundamental plane relation in a cluster at intermediate redshift.
They find that the fundamental plane relation of galaxies is very 
similar to that  of Coma, suggesting that the structure of the 
early-type galaxies has changed little since z$=$0.58.

\subsection{MS\thinspace 2137.3$-$2353}

MS2137$-$2353 is a very X-ray luminous cluster ($L_x=1.56\times 10^{45}$
erg\thinspace s$^{-1}$) at $z$=0.313 with a large, curved arc and the first
reported case of a radial gravitational image (Fort \etal 1992). The radial
image is embedded in the halo of the optically dominant galaxy and is best
seen in the HST image of Hammer \etal 1997.

The unique lensing configuration, with both a tangential and radial image 
and 3 arclets, allowed Mellier, Fort \& Kneib (1993) to develop a 
tightly constrained mass distribution model with only 2 free 
parameters (core radius and velocity 
dispersion) that fit the large arcs and predicted the positions of the 3 
arclets.  Their model assumed an elliptical cluster potential (circular
being ruled out due to the absence of a counter-arc). They found a
centrally-peaked mass distribution with a finite core radius that has 
a value of $r_c\simeq50\,h_{50}^{-1}$\thinspace kpc; considerably smaller
than the core radii derived from X-ray measurements, and in agreement
with the general trend that lensing clusters have centrally peaked mass 
density profiles. Allen (1998) shows that the strong gravitational 
lensing and X-ray mass  measurements for this cooling-flow cluster
are in excellent agreement, implying that the thermal pressure dominates 
over non-thermal processes in the support of the X-ray gas against
gravity in the central regions of the cooling-flow clusters, and 
also validating the hydrostatic assumptions used in the X-ray mass 
determination.

Hammer \etal 1997 studied in detail the core of this 
cluster using WFPC2 images. An analysis of the lensing properties
of the dark matter component indicates that within 30 to 150 h$_{50}^{-1}$
kpc from the mass center, the major axis and ellipticity of the dark matter
component are in rather good agreement with those derived from  X-ray 
and visible light, while the dark matter profile has a slope much flatter 
than that of the visible light (0.875 for the dark matter profile
vs 1.35 for the visible light profile).  MS2137$-$2353 is 
a good example of an essentially relaxed cluster, as is the case for
cooling-flow clusters, with an increasing mass-to-light ratio from 
the very center to \approx 150 h$_{50}^{-1}$ kpc.

\section{Arclet and candidate lensing clusters }

We believe that all 8 of the clusters in the
previous section show clear examples of lensing.  In this section,
the situation is not so obvious.  We present 8 additional clusters 
which contain structures we classify as arclets or candidate arcs.
Of these 8 clusters, we are confident that at least 2 of them 
(MS\thinspace 1008$-$12 and MS\thinspace 1231$+$15) are  true lens systems. 
The properties of the arclets and candidate arcs are listed in Table 3.

As mentioned earlier in \S4, we define an arclet simply as a short arc
with $l$\lthan 8$''$ and with an axis ratio 3\lthan$l/w$\lthan 10.  
Although, as with the giant arcs, we do not discriminate between 
high and low surface brightness arclets (the former sometimes called 
``mini-arcs''; e.g. see Lavery, Pierce \& McClure, 1993) for 
the most part our data  are not deep enough (see limits in Table 1) 
for us to have detected faint 
arclet populations similar to the ones in A1689 (Tyson, Valdes 
\& Wenk 1990; Tyson \& Fischer 1995) or  A2218 (Pello \etal 1992;
Kneib \etal 1995; Kneib \etal 1996).

Most of the arclets we find do not display any obvious curvature, partly
because they are short, but also possibly because of the presence of a
secondary deflector along the line of sight.  It is difficult, therefore,
to distinguish these putative arclets from blue, edge-on disk galaxies,
whether foreground, background, or cluster members, or possibly
interacting or morphologically-disturbed systems.
Although edge-on spirals often have well-defined nuclei, there are
examples where no nucleus can be seen, and without additional information
(e.g. spectra or infrared colors) these objects will be classified as
arclets.  However, we expect the number of cluster and projected edge-on 
spirals to be small (see discussion in \lefevre \etal 1994).
Furthermore, we expect the real arclets to be oriented more-or-less 
perpendicular to the cluster radius vector.
In the following subsections, we describe each arclet and candidate
lensing cluster in detail.

\begin{table*}[tb]
\begin{center}
Table 3. Candidate arcs and arclets
\vspace{.1in}
{
\small
\begin{tabular}{rcrrrrrrp{2.4in}}
\hline\hline
\multicolumn{1}{c}{(1)}&
\multicolumn{1}{c}{(2)}&
\multicolumn{1}{c}{(3)}&
\multicolumn{1}{c}{(4)}&
\multicolumn{1}{c}{(5)}&
\multicolumn{1}{c}{(6)}&
\multicolumn{1}{c}{(7)}&
\multicolumn{1}{c}{(8)}&
\multicolumn{1}{c}{(9)}\\
 & & & & & &\multicolumn{2}{c}{Photometry}& \\
\multicolumn{1}{c}{Cluster} &
\multicolumn{1}{c}{Arc} &
\multicolumn{1}{c}{$l$} & \multicolumn{1}{c}{$w$} &
\multicolumn{1}{c}{$l/w$} &
\multicolumn{1}{c}{$d_c$} &
\multicolumn{1}{c}{$B$} & \multicolumn{1}{c}{$R$} &
\multicolumn{1}{c}{References and Notes}\\
\hline
                           & &       &         &     &&        &      &\\
MS\thinspace 0451.5$+$0250 &1&5$''$.0&0$''$.9  &5.5  &21$''$&22.2    & 20.4 
&{\footnotesize \lefevre \etal (1994)} \\
MS\thinspace 0839.8$+$2938 &1&3.8    &$\cdots$ &$\cdots$&3&$\cdots$&$\cdots$  
&{\footnotesize radial arc?, buried in cD}\\
MS\thinspace 0906.5$+$1110 &1&$\cdots$&$\cdots$&$\cdots$&25& $\cdots$&23.6&\\
MS\thinspace 1008.1$-$1224 &1&6.0    &0.8      &7.5  &45&22.9    &21.1  
&{\footnotesize \lefevre \etal (1994)}\\
			   &2&6.0    &0.6      &10.0 &53&23.4    &22.1  &\\
MS\thinspace 1231.3$+$1542 &1&4.7    &0.9      &5.2  &13&$\cdots$&$V$=22.4 &
buried in halo\\
			   &2&5.3    &0.8      &6.6  &26&$\cdots$&$V$=23.2 &\\
MS\thinspace 1455.0$+$2232 &1&7.5    &1.3      &6.7  &21&$\cdots$&22.9  
&{\footnotesize \lefevre \etal (1994); Smail, Ellis \& Fitchett (1994); Smail \etal (1995)}\\
MS\thinspace 1546.8$+$1132 &1&4.2    &0.8      & 5.3 &35&23.1    &20.9  &\\
MS\thinspace 1910.5$+$6736 &1&6.1    &0.6      &10.5 &67&$\cdots$&20.6  
&{\footnotesize \lefevre \etal (1994)}\\
			   & &       &         &     &&        &      &\\
\hline
\end{tabular}
}
\end{center}
\end{table*}

\subsection{MS\thinspace 0451.5$+$0250}

This cluster lacks a well defined optical center.  The X-ray emission
is extended and may be associated with both the primary SW clump
of galaxies and a secondary clump located to the NE where we find
the candidate arclet.  The two clumps are separated by \approx 6$'$.  The image
of the candidate arclet shown in Figure 7 is the sum of the $B$ and $R$
frames.  The arclet is located 21$''$ to the NE of the large elliptical galaxy.
We cannot rule out that this is simply an 
edge-on cluster spiral.

\subsection{MS\thinspace 0839.8$+$2938}

The core of this rich, $z$=0.194 
cluster contains a number of short, blue, linear structures that
may be arclets or edge-on spirals.  The longest of these structures extends 
radially 3$''$.8 (16$\,h_{50}^{-1}$ kpc)  from the center of the cluster 
to the SE of the central cluster galaxy and may be 
another example of a radial arc (see Fig 7).  
We have also considered whether this structure could be a cooling
flow filament, since the brightest cluster galaxy has strong 
[OII], H$_\alpha$, and [NII] emission, that Nesci \etal (1989) interpreted 
as the signature of a large cooling flow.  Recent {\it ROSAT} observations 
have confirmed this cooling flow exists (Nesci, Perola \& Wolter 1995).
When the central particle density in a cooling flow
exceeds \approx 5--6$\times 10^{-3}$ cm$^{-3}$,
we expect to see strong optical emission lines (hydrogen recombination lines
and collisionally excited forbidden lines) from a $\sim$10 kpc 
asymmetrically-extended region around the cluster BCG (Hu, Cowie \&
Wang 1985; Cowie \etal 1983). 
Such filaments have been seen in the narrow band images of EMSS
cluster BCG's taken by DSG, so it is possible that the radial structure
in MS\thinspace 0839$+$2938 is such a filament. 
For now, we tentatively classify this radial structure as a candidate 
radial arc, and we intend to obtain a spectrum which will allow us to
distinguish between the lensing and cooling flow interpretations.

\subsection{MS\thinspace 0906.5$+$1110}

MS\thinspace 0906+1110 is a rich cluster with a large cD galaxy at $z$=0.180.
The X-ray luminosity is  $L_x=5.77\times10^{44}$ erg\thinspace s$^{-1}$.
The cluster displays obvious optical substructure with a 
bright elliptical galaxy \approx 95$''$ to the SW and another bright
elliptical nearly an equal distance to the NE, both along
the major axis of the cD.  These two galaxies are not visible in Fig. 7, 
but can be seen in the wide field image in Gioia \& Luppino (1994).
This cluster has no obvious giant arcs or bright arclets, but does contain
a number of faint arclets arranged tangentially around the cD.  
The largest of these is labeled in Figure 7 and is listed in Table 3.

\subsection{MS\thinspace 1008.1$-$1224}

MS\thinspace 1008$-$1224 is a very rich cluster at $z$=0.301 with
$L_x=4.49\times10^{44}$ erg\thinspace s$^{-1}$.  Yee {\it et al.}, 
1998, measured  112 cluster member redshifts and give an updated z$=$0.306.
This cluster contains at least one obvious example of a lensed arclet, 
but no giant arcs (\lefevre \etal 1994).  When displayed
at high contrast, the cluster image exhibits a clear patterned circularity
centered on the brightest cluster galaxy.  The longest arclet located 51$''$ 
to the North of the  BCG is obviously curved.  There is another 
bright, blue candidate arclet to the East of the BCG (see Figure 8).

\subsection{MS\thinspace 1231.3$+$1542}

This $z$=0.238 cluster contains a faint arc 4$''$.7 long located midway
between the two brightest cluster galaxies, a bright arclet 25$''$.7 to the 
south of the BCG, and a faint arclet to the North of the second ranked 
galaxy.  The faint arc and arclet are difficult to see in Figure 8
because they are  embedded in the common halo of the cluster dumbbell 
galaxies. This cluster and lens system is similar to MS\thinspace 0302+1658.
Both clusters have a lensed arc or arclet midway between two 
nearly-equal-brightness dumbbell galaxies in the cluster core.  With 
$L_x=2.88\times10^{44}$  erg\thinspace s$^{-1}$, this cluster is one of 
the lowest $L_x$ EMSS clusters to display obvious lensing.

\subsection{MS\thinspace 1455.0$+$2232}

This cluster was first discovered by Schild \etal (1980) and later 
observed by Mason \etal (1981) who measured the redshift of $z$=0.259.  
MS\thinspace 1455$+$2232 is among  the most X-ray luminous
clusters in the EMSS with $L_x=1.6\times10^{45}$ erg\thinspace s$^{-1}$.
It also has, by far, the largest narrow-band line emission luminosity
in the cooling flow survey of DSG (see Table 1).

MS\thinspace 1455+22 has been the focus of a weak lensing study by
Smail, Ellis \& Fitchett  (1994) and Smail {\it et al.} (1995) 
using the weak lensing technique of Kaiser
\& Squires (1993).  Smail \etal 1994 detect a clear lensing signal from
the cluster and estimate a lower limit to the projected mass within the 
central \approx 0.9 Mpc of \approx 2.2$\times 10^{14}$ $M_{\sun}$.
However, their images only had a field of view of $5'\times5'$
and so only cover the core of the cluster and do not provide
any information about the mass distribution at large radii.
Smail \etal 1995 have also presented {\it ROSAT} HRI data that detect extended
emission out to $\sim$1 Mpc and resolve the central 50\thinspace kpc.
Their data show the X-ray emission is highly peaked on the cD galaxy 
and confirm the cluster has a massive
cooling flow of 630$^{+257}_{-178}\; M_{\sun}$ yr$^{-1}$.

This cluster contains a \approx 7$''$.5 long 
arc-like structure located 20$''$ NE of the BCG (see Figure 7). 
This putative arc is also visible in Smail \etal images,  which are 
considerably deeper than ours (for this particular cluster) and have 
comparable seeing. 
Upon close inspection, we notice what may be a galaxy nucleus to one side 
and the arc appears slightly curved, but in the direction away from the
central galaxy.  Although we include this arclet in our list of candidates, 
we consider this a questionable case.

\subsection{MS\thinspace 1546.8$+$1132}

MS\thinspace 1546$+$11 is a rich, elongated cluster with a large,
double-nucleus cD galaxy.  The ellipticity of the cD halo is
oriented in the same direction (EW) as the 
elliptical distribution of the optical galaxies, and
orthogonal to the 2 arc-like structures to the East of the cD.
The largest candidate arc is 4$''$.2 long, is bright, and appears
to be marginally resolved.  
Again, this could simply be an edge-on cluster galaxy since it has
a somewhat lenticular shape, but we see no obvious bright nucleus.

\subsection{MS\thinspace 1910.5$+$6736}

This cluster at $z$=0.246 has a double nucleus BCG and is rather poor
with an irregular morphology.  There are 2 elongated, linear structures 
located \approx 1$'$ to the SE of the BCG. 
The northernmost object has an obvious core and looks like an edge-on spiral.  
The other object, however, is \approx 6$''$ long, has
no apparent core, and appears unresolved.  \lefevre \etal (1994) classified
this object as a giant luminous arc since it had an axis ratio $l/w>10$
and was oriented perpendicular to the radius vector to the BCG.  
Our criteria in this paper, however, require that an object have $l\geq 8''$
to be called a giant arc. Furthermore, we believe the lensing explanation for
this object is less convincing than in some of the other cases, particularly
since the object has no apparent curvature.  

\section{Discussion}

From inspection of Table 1, the lensing clusters do not peak around
any preferred redshift, as might be expected if clusters are evolving with
redshift (Bartelmann 1993; Wu 1993). It is apparent that although we  have
a substantial number of clusters at high redshift (5 with $z$\gthan 0.5), 
completeness may be a problem beyond  $z$\approx 0.35$\,-\,$0.4.
The apparent increase in the lensing fraction 
with redshift is no doubt a selection bias, resulting from the fact that  
we can only see the most X-ray luminous clusters at the high redshifts,
and these very luminous clusters are likely to be more massive,
thus increasing our chances of finding lenses. This effect is illustrated
by the  dramatic increase in the lensing fraction with increasing 
$L_x$. Three of the 5 EMSS
clusters with $L_x$\gthan 10$^{45}$ erg\thinspace s$^{-1}$, and none of the 15 
EMSS clusters with $L_x$\lthan 4$\times$10$^{44}$ erg\thinspace s$^{-1}$ contain 
giant arcs. The giant arc lensing fraction for clusters having 
$L_x$\gthan 4$\times$10$^{44}$ erg\thinspace s$^{-1}$ is 30\%. Although the 
statistics are poor because the number of high-$L_x$ clusters is small, 
it appears that high X-ray luminosity does indeed point to the most
massive clusters. We find no evidence of lensing in the 2 clusters
beyond $z$\approx 0.7. One could interpret this as a real cutoff,
thus constraining the redshift distribution of the background sources.
The more likely explanation, however, is that our images for these
clusters are simply not deep enough. For clusters beyond $z$\approx 0.7,
the sources will be at $z$\gthan 1-1.5 and consequently will be very faint
with only the most luminous ($>>L^\ast$) galaxies visible. 
Deeper images will reveal faint arcs, if they exist, in these clusters as is 
the case of MS\thinspace 1137+6625, at z$=$0.78,  where
a system  of giant arcs present in the core of the cluster
is revealed by a 8700 s long exposure in R-band taken with 
the Keck II telescope (see Fig. 1 in Clowe \etal 1998).

A diagnostic of the mass density profile is provided by the
arc widths (Hammer 1991). The observed width ($w_{\rm arc}$) of a 
lensed arc is related to the intrinsic
source width ($w_s$) through the following relation,
$w_{\rm arc}={1\over 2}\,w_s\,(1-K_{\rm arc})^{-1}$.
All of the information on the mass density profile is 
contained in the so-called matter term $K_{\rm arc}$ which can vary 
from 0 to 0.5 for compact, singular point mass, singular isothermal 
sphere or $r^{1/4}$ profiles, or from 0.5 to 1 for a 
non-singular isothermal sphere.  Therefore,  if we assume that 
the arcs are drawn from a population of field galaxies 
whose intrinsic widths are more-or-less constant, then we see 
that more compact lenses produce  thinner  arcs. 
The majority of the arcs in this sample are thin, often unresolved or
marginally resolved, even in 0$''$.5$\,-\,$0$''$.7 seeing. Thus, we
find that the majority of the clusters must have compact cores. 
Obviously, the assumptions about the field galaxy sizes are critical
in this interpretation. Hammer (1991) and Hammer
\etal (1993) argue that faint field galaxies are in general resolved
or marginally resolved with intrinsic FWHM of order 0$''$.5$\,-\,$1$''$
(Lilly, Cowie \& Gardner 1991; Tresse \etal 1993; Colless \etal 1994),
in agreement with our conclusions.

In addition to the arc widths, the arc locations, curvatures
and orientations also depend on the mass density profile.
The arc radius of curvature can be considered to be an upper limit to
the cluster core radius (Bartelmann, Steinmetz \& Weiss 1995). Furthermore,  
\miralda (1993a) and Grossman \& Saha (1994) show that when arcs
trace ellipses whose major axes are aligned with the
cluster major axis (i.e. the arcs are perpendicular to the cluster
major axis), then the cluster mass density profile must be
steeper than isothermal.  If arcs trace ellipses aligned with the 
cluster minor axis, then the mass density profile is shallower than 
isothermal.

When we examine the giant arcs in our sample, we find none that are
perpendicular to the cluster minor axis.  In four of the giant arc clusters,
the optical galaxy distribution is clearly elongated, and in these
four clusters, the arcs are orthogonal to a radius vector that is more-or-less 
aligned with the optical (and in MS\thinspace 0451$-$0305,
the X-ray as well) major axis. The other four giant arc clusters have 
no obvious optical axis of symmetry. Similarly, the arclets and candidate 
arcs also tend to
be normal to the cluster major axis, in the cases where a cluster
has a clear axis of symmetry. From this evidence, we conclude that in 
most our examples of lensing, the mass density profiles appear to 
be steeper than isothermal. Of course, in arriving at this conclusion, 
we are assuming that the optical galaxies are tracing the total cluster 
mass. We also point out, however, that several of the clusters  
(e.g. MS\thinspace 2137$-$23 and MS\thinspace 0839+29) may contain radial 
gravitational images, which requires a cluster density profile with a finite 
core radius (e.g. Grossman \& Saha 1994; Mellier, Fort \& Kneib 1993). 

\section{Summary and Conclusions}

We have presented the results of our systematic  search for lensed arcs 
and arclets in a large, homogeneous, X-ray-selected sample of 38 distant 
clusters of galaxies.   Below, we summarize our conclusions.

\begin{enumerate}
\item{
  Our survey has yielded 8 clusters with giant arcs,
  2 clusters with arclets, and 6 candidate lensing systems. 
  }
\item{
  Our results indicate that high X-ray luminosity 
  ($L_x$\gthan 4$\times$10$^{44}$ erg\thinspace s$^{-1}$) does indeed  
  identify the most massive clusters, and thus X-ray selection is the 
  preferred method for finding true, rich clusters at intermediate and 
  high redshifts.
  }
\item{The majority of the arcs in our sample have large axis ratios, 
  are thin, and in those clusters with obvious optical axes of symmetry, 
  the arcs tend to be oriented orthogonal to the optical major axis. 
  Furthermore, the high lensing frequency in our sample (21\% giant arcs in 
  the entire sample, 30\% giant arcs for clusters with $L_x$\gthan 4$\times$10$^{44}$ 
  erg\thinspace s$^{-1}$, 60\% for clusters with $L_x$\gthan 1$\times$10$^{45}$ 
  erg\thinspace s$^{-1}$) is consistent with the lensing frequency  predicted 
  by statistical models that assume clusters have compact cores.

  However, our conclusion that the central regions of these clusters have 
  compact mass density profiles does not preclude the presence of large,
  extended dark matter halos that is convincingly demonstrated 
  to exist through the detection of weak lensing at large 
  radii (Smail, Ellis, \&  Fitchett, 1994; Kaiser, Squires \& Broadhurst 1996; 
  Squires \etal 1997; Geiger \& Schneider, 1998; Clowe \etal 1998; 
  Kaiser \etal 1998).
}
\end{enumerate}

Future observations and theoretical modelling will provide important
information on these cluster lenses.  We are presently acquiring optical
spectra of the clusters and arcs with the CFHT and Keck telescopes in order
to measure the arc redshifts and the cluster velocity dispersions.
These spectra will allow us to confirm the lensing hypothesis, epecially
for the candidate arcs and arclets, and will be crucial for theoretical
modelling of the more interesting lensing configurations.  

\begin{acknowledgements}

We thank the UH Time Allocation Committee for their generous allocation of 
UH 2.2\thinspace m and CFHT time for this project.  We also thank
John Stocke and Simon Morris for sharing their data, particularly the
MS0451$-$03 redshift, prior to publication.  We have enjoyed fruitful
discussions and interactions with Megan Donahue, Anna Wolter, Pat Henry, 
Neal Trentham, and with our CNOC colleagues (E. Ellingson, H. Yee, 
R. Carlberg, R. Abraham) and others (N. Kaiser, Y. Mellier, B. Fort). 
We are grateful to all those (P. Henry, B. Tully,
L. Kofman, J. Dalcanton, L. Cowie, M. Metzger, S. Miyazaki)
who read and commented on early drafts of this paper, and to Doug
Clowe, who helped with preliminary FOCAS analysis of some of the 
clusters. Finally, we appreciate the patience and generosity 
of Brent Tully who allowed us to ``take an image or two'' during 
several of his observing runs.
The UH CCD cameras were constructed using NSF Grant AST-9020680.
This work also received partial financial support from NASA-STScI grant
GO-05987.02-94A, NSF Grants AST-9119216 and AST-9500515, 
NASA Grants NAG5-2594, NAG5-2914, and ASI grants ARS-94-10 and ARS-96-13.

\end{acknowledgements}

\begin{figure*}[tbhp]
\vspace{2 cm}
\vspace{-1.0truein}
\par\noindent
\begin{center}
\caption{R-band images for seven of the eight
EMSS clusters with giant arcs.
}
\end{center}
\end{figure*}

\begin{figure*}[tbhp]
\vspace{2 cm}
\vspace{-1.0truein}
\par\noindent
\begin{center}
\caption{True color image of MS\thinspace 0302$+$1658
showing the blue giant arc system.  The image is a 512 $\times$ 512 pixel
subarray (750 kpc $\times$ 750 kpc
at $z$=0.426; $H_0$\equals 50, $q_0\,$=$\,{1\over 2}$)
that was produced using the $V$ (5400\thinspace s), $R$ (6000\thinspace s),
and $I$ (3000\thinspace s) CCD frames. North is up and East is to the left.
}
\end{center}
\end{figure*}

\begin{figure*}[tbhp]
\vspace{2 cm}
\vspace{-1.0truein}
\par\noindent
\begin{center}
\caption{True color image of MS0451$-$03 produced using
the $V$ (8400\thinspace s), $R$ (7200\thinspace s),
and $I$ (4200\thinspace s) CCD frames.  The image is a 750 $\times$ 750 pixel
subarray and measures $2'.75 \times 2'.75$ (1.2 Mpc $\times$
1.2 Mpc at $z$=0.55 for $H_0$\equals 50, $q_0\,$=$\,{1\over 2}$).
North is up and East is to the left.
}
\end{center}
\end{figure*}

\begin{figure*}[tbhp]
\vspace{2 cm}
\par\noindent
\vspace{-1.0truein}
\begin{center}
\caption{
A high contrast image of the core of MS\thinspace 0451.6$-$0305 at
$z$=0.55.  The giant arc A1 and bright arclet A2 are clearly visible.
Note that arc A1 has obvious structure along its length, and
little if any curvature.}
\end{center}
\end{figure*}

\begin{figure*}[tbhp]
\vspace{2 cm}
\vspace{-1.0truein}
\begin{center}
\par\noindent
\caption{R-band image of MS\thinspace 1358.4$+$6245. The lower panel reveals 
the faint, curved arc to the SW of the central galaxies, corresponding to the
4.92 red arc in the HST images of Franx \etal 1997.
The axes are in arcsec, and the scale at the cluster is 
5.8$\,h_{50}^{-1}$ kpc/arcsec.
}
\end{center}
\end{figure*}

\begin{figure*}[tbph]
\vspace{2 cm}
\vspace{-1.0truein}
\begin{center}
\caption{R-band images of EMSS clusters with arclets and candidate arcs.
}
\end{center}
\end{figure*}

\begin{figure*}[tbph]
\vspace{2 cm}
\vspace{-1.0truein}
\begin{center}
\caption{R-band images of EMSS clusters with arclets and candidate arcs.
}
\end{center}
\end{figure*}


\begin{thebibliography}{}

\bibitem{} Allen, S.W., 1998, MNRAS, 296, 392
\bibitem{} Annis, J. 1994, PhD Thesis, University of Hawaii.
\bibitem{} Bartelmann, M. 1993, A\&A, 276, 9.
\bibitem{} Bartelmann, M., Steinmetz, M. \& Weiss, A. 1995, A\&A, 297, 1. 
\bibitem{} Bautz, M., Honda, E., Ventrella, J. \& Gendreau, K.,
	1997, on ``X-ray Imaging and Spectroscopy of Cosmic Hot Plasma'', 
	Proceedings of the International Symposium on X-ray astronomy, 
	Universla Academy Press Publ., Eds. Makino, F. and Mitsuda, K., p. 75.
\bibitem{} Broadhurst, T., Taylor, A., \& Peacock, J. 1995, ApJ, 438, 49.
\bibitem{} Carlberg, R., Yee, H., Ellingson, E., Abraham, R., Gravel, P., 
	Morris, S. \& Pritchet, C.J., 1996, ApJ, 462, 32.
\bibitem{} Carlberg, R., Yee, H., \& Ellingson, E., 1997, ApJ, 478, 462.
\bibitem{} Carlberg, R.G., Yee, H.K.C., Ellingson, E., Morris, S.L., Abraham, R.,
	Gravel, P., Pritchet, C., Smecker-Hane, T.,  Hartwick, F.,
	Hutchings, J., Oke, J.B. (the CNOC cluster collaboration) 1997a  
	ApJL, 485, L13.
\bibitem{} Carlberg, R.G., Morris, S.L., Yee, H., \& Ellingson, E., 1997b
	ApJL, 479, L19.
\bibitem{} Carlberg, R., \etal, 1998, in ``Large Scale Structure: Tracks 
	and Traces'', Proceedings of the 12th Potsdam Cosmology Workshop,
	1997, Eds. V. Mueller, S. Gottloeber, J.P. Muecket, J. Wambsganss,
	World Scientific 1998, p. 119-126.
\bibitem{} Castander, F., Ellis, R., Frenk, C., Dressler, A., \& Gunn, J.
	1994, ApJL, 424, L79.
\bibitem{} Colless, M., Ellis, R., Taylor, K. \& Hook, R., 1990, MNRAS, 
	244, 408.
\bibitem{} Colless, M., Ellis, R., Broadhurst, T., Taylor, K., and
	Peterson, B. 1993, MNRAS, 261, 19.
\bibitem{} Colless, M., Schade, D., Broadhurst, T. \& Ellis, R. 1994,
	MNRAS, 267, 1108.
\bibitem{} Clowe, D.I., 1998, PhD thesis, University of Hawaii.
\bibitem{} Clowe, D.I., Luppino, G., Kaiser, N., Henry, J.P. \&  
         Gioia, I.M., 1998, ApJL, 497, L61.
\bibitem{} Colless, M.M., Ellis, R.S., Taylor, K. \& Hook, R.N., 1990,
	MNRAS, 244, 408.
\bibitem{} Colless, M.M., Ellis, R.S., Broadhurst, T., Taylor, K. \& 
	Peterson, B.A., 1993, MNRAS, 261, 19.
\bibitem{} Collins, C. A., Burke, D. J., Romer, A. K., Sharples, R. M., \& 
	Nichol, R. C. 1997, ApJL, 479, L117.
\bibitem{} Cowie, L., Hu, E., Jenkins, E., York, D. 1983, ApJ, 262, 29.
\bibitem{} Donahue, M., Stocke, J., \& Gioia, I. 1992, ApJ, 385, 49.
\bibitem{} Donahue, M. \& Stocke, J. 1995, ApJ, 449, 554.
\bibitem{} Donahue, M.E., 1996, ApJ, 468, 79.
\bibitem{} Donahue, M.E., Voit, M., Gioia, I.M., Luppino, G.A., Hughes,
	J.H. \& Stocke, J.T., 1998, ApJ, 552, 550.
\bibitem{} Dressler, A. \& Gunn, J. 1992, ApJS, 78, 1.
\bibitem{} Ebeling, H., Edge, A.C., Fabian, A.C., Allen, S.W., Crawford, C.
	S. \& Boheringer, H., 1997, ApJL, 479, L101.
\bibitem{} Edge, A.C., Stewart, G.C. \& Fabian, A.C., 1992, MNRAS, 258, 157.
\bibitem{} Ellingson, E., Yee, H. K. C., Abraham, R. G., Morris, S. L., 
	Carlberg, R. G. \& Smecker-Hane, T. 1997, ApJS, 113, 1.
\bibitem{} Ellingson, E., Yee, H.K.C.,  Abraham, R., Morris, S.L. \& 
	Carlberg, R.G., 1998, ApJS, 116, 247.
\bibitem{} Fabricant, D., McClintock, J. \& Bautz, M. 1991, ApJ, 381, 33.
\bibitem{} Fabricant, D., Bautz, M. \& McClintock, J. 1994, AJ, 107, 8.
\bibitem{} Fahlman, G., Kaiser, N., Squires, G. \& Woods, D. 1994, ApJ,
	437, 56.
\bibitem{} Fisher, D., Fabricant, D., Franx, M. \& van Dokkum, P., 1998,
	ApJ, 498, 195.
\bibitem{} Fort, B., LeFevre, O., Hammer, F. \& Cailloux, M. 1992, ApJL, 399,
	L125.
\bibitem{} Franx, Marijn, Illingworth, G.D., Kelson, D.D., van Dokkum, P.G. 
	\& Tran, K.V.Y, 1997, ApJL, 486, L75.
\bibitem{} Geiger, B. \& Schneider, P., 1998, MNRAS, 295, 497.
\bibitem{} Gioia, I., Feigelson, E., Maccacaro, T., Schild, R. \& Zamorani,
	G. 1983, ApJ, 271, 524.
\bibitem{} Gioia, I. M., Maccacaro, T., Schild, R.E., Wolter, A., Stocke,
	J.T., Morris, S. L. \&  Henry, J. P. 1990a, ApJS, 72, 567.
\bibitem{} Gioia, I.M., Henry, J. P., Maccacaro, T., Morris, S., Stocke, J.,
	   \& Wolter, A. 1990b, ApJL, 356, L35.
\bibitem{} Gioia, I. \& Luppino, G. A. 1994, ApJS, 94, 583.
\bibitem{} Gioia, I.M., Shaya, E.,  Le\thinspace F\`evre, O., Falco, E.E., 
	Luppino, G.A. \& Hammer, F., 1998a, ApJ, 497, 573.
\bibitem{} Gioia, I.M., Henry, J. P., Mullis, C.R., Ebeling, H. \& Wolter, 
	A., 1998b, AJ, in press.
\bibitem{} Giraud, E. 1991, ESO Messenger, 66.
\bibitem{} Giraud, E. 1992, A\&A 259, L49.
\bibitem{} Grossman, S. \& Saha, P. 1994, ApJ, 431, 74
\bibitem{} Hammer, F. 1991, ApJ, 383, 66.
\bibitem{} Hammer, F., Angonin-Willaime, M.C., Le\thinspace F\`evre, O., Wu, 
	X.P., Luppino, G.A. \& Gioia, I.M. 1993, in {\it Gravitational 
	Lenses in the Universe}, 31st Liege Int. Astroph. Colloquium, 
	Universit\'e de Li\`ege, eds. J. Surdej, D. Fraipont-Caro, E. Gosset, 
	S. Refsdal and M. Remy, 609.
\bibitem{} Hammer, F., Gioia, I.M., Shaya, E.J., Teyssandier, P., 
	Le\thinspace F\`evre, O. \&  Luppino, G.A.,  1997, ApJ, 491, 477.
\bibitem{} Henry, J.P., 1997, ApJL, 489, L1. 
\bibitem{} Henry, J. P., Gioia, I. M., Maccacaro, T., Morris, S. L.,
	Stocke, J. T. \& Wolter, A. 1992, ApJ, 386, 408.
\bibitem{} Henry, J.P., Gioia, I.M., Huchra, J.P., Burg, R. McLean, B.,
	Boehringer, H., Bower, R.G., Briel, U.G., Voges, W., MacGillivray, H. 
	\& Cruddace, R.G., 1995, ApJ, 449, 422.
\bibitem{} Hoekstra, H., Franx, M., Kuijken, K. \& Squires, G., 1998,
	ApJ, 504, 636.
\bibitem{} Hu, E., Cowie, L. \& Wang, Z. 1985, ApJS, 59, 447.
\bibitem{} Kaiser, N., 1992, ApJ, 388, 272.
\bibitem{} Kaiser, N. \&  Squires, G. 1993, ApJ, 404, 441.
\bibitem{} Kaiser, N., Squires, G. \& Broadhurst, T., 1996, ApJ, 449, 460.
\bibitem{} Kaiser, N., Wilson, G., Luppino, G., Kofman, L., Gioia, I.M., 
	Metzger, M. \& Dahle, H., 1998, ApJ, submitted, astro-ph/9809268.
\bibitem{} Kelson, D.D., van Dokkum, P.G., Franx, M. \& Fabricant, D.,
	1997, ApJL, 478, L13.
\bibitem{} Kneib, J.P., Mellier, Y., Pello, R., Miralda-Escud\`e, J.,
	Le Borgne, J.-F., Bohringer, H. \& Picat, J.-P., 1995, A\&A, 303, 27.
\bibitem{} Kneib, J.P., Ellis, R.S., Smail, I.,  Couch, W.J. \& Sharples, R.M.,
	1996, ApJ, 471, 643.
\bibitem{} Koo, D.C.  \& Kron, R.G., 1992, ARA\&A, 30, 613.
\bibitem{} Kovner, I.  1987, ApJ, 321, 686.
\bibitem{} Lavery, R., Pierce, M., and McClure, R. 1993, ApJ, 418, 43.
\bibitem{} Le\thinspace F\`evre, O., Hammer, F., Angonin, M.C., Gioia, I.M. \& 
	Luppino, G. A., 1994, ApJL, 422, L5.
\bibitem{} Lilly, S.J., Cowie, L. \&  Gardner, 1991, ApJL, 369, L79.
\bibitem{} Lilly, S.J., Tresse, L., Hammer, F., Crampton, D. \& 
	Le\thinspace F\`evre, O., 1995, ApJ, 455, 108.
\bibitem{} Luppino, G., Cooke, B., McHardy, I., and Ricker, G. 1991, AJ 102, 1.
\bibitem{} Luppino, G. A. \&  Gioia, I. M. 1992, A\&A, 265, L9.
\bibitem{} Luppino, G.,  Gioia, I., Annis, J., Le\thinspace F\`evre, O., 
           \&  Hammer, F., 1993, ApJ, 416, 444.
\bibitem{} Luppino, G.A. \&   Gioia, I.M. 1995, ApJL, 445, L77.
\bibitem{} Luppino, G.A. \&  Kaiser, N., 1997, ApJL, 475, L20
\bibitem{} Lynds, R. \& Petrosian, V. 1989, ApJ, 336, 1. 
\bibitem{} Maccacaro, T., Wolter, A., McLean, B., Gioia, I., Stocke, J., 
	Della Ceca, R., Burg, R. \&  Faccini, R. 1994, 
	Astrophys. Letters and Comm., Gordon \& Breach Pubs., 29, N. 5-6, 267.
\bibitem{} Mason, K., Spinrad, H., Bowyer, S., Reichert, G.,
	\&  Stauffer, J. 1981, AJ, 86, 803.
\bibitem{} Mathez, G., Fort, B., Mellier, Y., Picat, J.-P. \&  Soucail, G.
	 1992, A\&A, 256, 343.
\bibitem{} McClure, R., Grundmann, W., Rambold, W., Fletcher, J.,
	Richardson, H., Stilburn, J., Racine, R., Christian, C., 
	\&  Waddell, P., 1989, PASP, 101, 1156.
\bibitem{} Mellier, Y., Fort, B. \&  Kneib, J.-P. 1993, ApJ, 407, 33.
\bibitem{} Mellier, Y., Fort, B., Bonnet, H. \&  Kneib, J.-P. 1994, in
	{\it Cosmological Aspects of X-ray Clusters of Galaxies}, 
	NATO Advanced Study Institute, ed W.C. Seitter, 219.
\bibitem{} Miralda-Escud\`e~, J., 1991, ApJ, 370, 1.
\bibitem{} Miralda-Escud\`e~, J. 1992, ApJL, 390, L65.
\bibitem{} Miralda-Escud\`e~, J. 1993a, ApJ, 403, 497.
\bibitem{} Miralda-Escud\`e~, J. 1993b, ApJ, 403, 509.
\bibitem{} Mullis, C.R., Gioia, I.M. \&  Henry, J.P. 1998,
	in IAU Symposium 188 ``The Hot Universe'', Kyoto, 1997, Aug 26-30.
\bibitem{} Navarro, J.F., Frenk, C.S. and White, S.D.M., 1996, ApJ, 462, 563.
\bibitem{} Nesci, R., Gioia, I., Maccacaro, T., Morris, S., Perola, G.,
	Schild, R. \&  Wolter, A. 1989, ApJ, 344, 104.
\bibitem{} Nesci, R., Perola, G. \&  Wolter, A. 1995, A\&A, 299, 34.
\bibitem{} Pello, R., LeBorgne, J.-F., Soucail, G., Mellier, Y. \& 
	Sanahuja, B. 1991, ApJ, 366, 405.
\bibitem{} Pello, R., LeBorgne, J., Sanahuja, B., Mathez, G. \& Fort, B.
	1992, A\&A, 266, 6.
\bibitem{} Pesce, J.E., Fabian, A. C., Edge, A. C. \&  Johnstone, R. M. 1990,
           MNRAS, 244, 58.
\bibitem{} Piccinotti, G., Mushotzky, R., Boldt, E., Holt, S., Marshall, F.,
	Serlemitsos, P. \&  Shafer, R. 1982, ApJ, 253, 485.
\bibitem{} Pierre, M., Le Borgne, J.J., Soucail, G. \& Kneib, J.P., 1996, A\&A, 311, 413.
\bibitem{} Rector, T.A., Stocke, J.T. \&  Perlman, E., 1998, ApJ, submitted
\bibitem{}Rosati, P., Della Ceca, R., Norman, Colin \&  Giacconi, R., 1995, 
	ApJL, 445, L11
\bibitem{} Rosati, P., Della Ceca, R., Norman, Colin \&  Giacconi, R., 1998,
	ApJL, 492, L21
\bibitem{} Scharf, C.A., Jones, L.R.L., Ebeling, H., Perlman, E., 
	Malkam , M. \&  Wegner, G. 1997, ApJ, 477, 79.
\bibitem{} Schild, R., Leach, R., Weekes, T. \&  Gursky, H. 1980, AJ, 85, 121.
\bibitem{} Schneider, P., 1996, MNRAS, 283, 837.
\bibitem{} Seitz, S. \&  Schneider, P. 1996, A\&A, 305, 383.
\bibitem{} Seitz, S. \&  Schneider, P. 1998, astro-ph/9802051.
\bibitem{} Smail, I., Ellis, R. S., Fitchett, M. J., Norgaard-Nielsen, H. U., 
	   Hansen, L. \&  Jorgensen, H. E. 1991, MNRAS, 252, 19.
\bibitem{} Smail, I., Ellis, R. \&  Fitchett, M. 1994, MNRAS, 270, 245.
\bibitem{} Smail, I., Ellis, R., Fitchett, M. \&  Edge, A. 1995, MNRAS, 
	273, 277.
\bibitem{} Squires, G., Kaiser, N., Fahlman, G., Babul, A., Woods, D.,
	Neumann, D.M. \&  Bohringer, H., 1996a, ApJ, 461, 572.
\bibitem{} Squires, G., Kaiser, N., Fahlman, G., Babul, A. \&  Woods, D.,
 	1996b, ApJ, 469, 73.
\bibitem{} Squires, G., Neumann, D.M., Arnaud, M., Babul, A., Bohringer, H.
	Fahlman, G. \&  Woods, D., 1997, ApJ, 482, 648.
\bibitem{} Stocke, J.T., Morris, S.L., Gioia, I.M., Maccacaro, T., Schild,
	   R.E., Wolter, A., Fleming, T.A. \&  Henry, J.P. 1991, ApJS, 76, 813.
\bibitem{} Tresse, L., Hammer, H., Le\thinspace F\`evre, O. \& 
           Proust, D. 1993, A\&A, 277, 53.
\bibitem{} Tyson, J. A., 1988, Nature, 334, 294.
\bibitem{} Tyson, J.A., Valdes, F. \& Wenk, R.A., 1990, ApJL, 349, L1.
\bibitem{} Tyson, J.A. \& Fischer, P., 1995, ApJ, 446, 55.
\bibitem{}  van Dokkum, P.G., Franx, M., Illingworth, G.D., Kelson, D.D., 
	Fisher, D. \& Fabricant, D., 1998, ApJ, 500, 714.
\bibitem{} Vikhlinin, A., McNamara, B.R., Forman, W., Jones, C., Quintana, H.
	\& Hornstrup, A., 1998a, ApJ, 502, 558.
\bibitem{} Vikhlinin, A., McNamara, B.R., Forman, W., Jones, C., Quintana, H.
	\& Hornstrup, A., 1998b, ApJL, 498, L21.
\bibitem{} Wu, X. P. \& Hammer, F. 1993, MNRAS, 262, 187.
\bibitem{} Wu, X. P. 1993, A\&A, 270, L1.
\bibitem{} Yee, H. K. C., Ellingson, E., Morris, S.L., Abraham, R. \&
	Carlberg, R. G., 1998, ApJS, 116, 211.
\end{thebibliography}
\end{document}